\documentclass[pra,a4paper,twocolumn,superscriptaddress,pra,eqsecnum]{revtex4}

\usepackage{amsfonts,amsmath,graphicx,color,epsfig,amssymb}
\DeclareGraphicsExtensions{.jpeg,.jpg,.ps,,.pdf}

\definecolor{nblue}{rgb}{0.2,0.2,0.7}
\definecolor{ngreen}{rgb}{0.1,0.5,0.1}
\definecolor{nred}{rgb}{0.7,0.2,0.2}

\newcommand{\ea}{{\it et al.~}}

\newcommand{\beq}{\begin{equation}}
\newcommand{\eeq}{\end{equation}}
\newcommand{\bqa}{\begin{eqnarray}}
\newcommand{\eqa}{\end{eqnarray}}
\newcommand{\erf}[1]{Eq.~(\ref{#1})}

\newcommand{\bra}[1]{\langle{#1}|}
\newcommand{\ket}[1]{|{#1}\rangle}
\newcommand{\sch}{Schr\"odinger}

\newcommand{\sq}[1]{\left[ {#1} \right]}

\newcommand{\scu}[1]{\{{#1}\}}
\newcommand{\ro}[1]{\left( {#1}\right)}

\newcommand{\san}[1]{\langle{#1}\rangle}

\newcommand{\tr}[1]{\mathrm{Tr}\sq{ {#1}}}

\newcommand{\s}[1]{\hat\sigma_{#1}}

\begin{document}

\title{Nonlocality of a single photon: paths to an EPR-steering experiment}
 \author{S. J. Jones}
\affiliation{ Centre for Quantum Computation and Communication Technology (Australian Research Council),   Centre for Quantum Dynamics, Griffith University, Brisbane, 4111, Australia }
  \author{H. M. Wiseman}
\affiliation{ Centre for Quantum Computation and Communication Technology (Australian Research Council),   Centre for Quantum Dynamics, Griffith University, Brisbane, 4111, Australia }

    \date{\today}

\begin{abstract}
A single photon 
incident on a beam splitter produces an entangled field state, and in principle
could be used to violate a Bell-inequality, but such an experiment (without post-selection)
is beyond the reach of current experiments. Here we consider the somewhat
simpler task of demonstrating EPR-steering with a single photon (also without post-selection).
That is, of demonstrating that Alice's choice of measurement on her ``half'' of a single photon
can affect the other ``half'' of the photon in Bob's lab, in a sense rigorously defined
by us and Doherty [Phys. Rev. Lett. {\bf 98}, 140402 (2007)]. Previous work by Lvovsky and co-workers [Phys. Rev. Lett. {\bf 92}, 047903 (2004)]
has addressed this phenomenon (which they called ``remote preparation'') experimentally using homodyne
measurements on a single photon.
Here we show that, unfortunately, their experimental parameters do not meet the bounds necessary for a rigorous demonstration of EPR-steering
with a single photon. However, we also show that modest improvements in the experimental
parameters, and the addition of photon counting to the arsenal of Alice's measurements,
would be sufficient to allow such a demonstration.
   \end{abstract}

\maketitle

\section{Introduction}

The nonlocal properties of a single photon or particle is of continuing
interest, both theoretically  \cite{TanWalColPRL91, HarPRL94, WisVac03, EnkPRA05, DunVedPRL07}
and experimentally \cite{LeeKimPRA01,LomEtalPRL02,BabRieLvoEPL03,BabBreLvoPRL04,BabEtalPRL04,HesEtalPRL04}. 
It is well known, and experimentally verified \cite{LomEtalPRL02,BabEtalPRL04},  
that splitting a single photon using a beam splitter produces an
entangled field state. Its entanglement is  ``accessible''  \cite{BarWisPRL03} or ``extractable'' \cite{JonEtalPRA06}   providing there exists other fields
which can be interfered with one or both ``halves'' of the single photon prior to
detection (as photo-detection itself is a phase-insensitive operation).
In particular, strong local oscillators (LOs) were used to perform the homodyne tomography
which verified entanglement in  Ref.~\cite{BabEtalPRL04},  while it was shown theoretically that
weak LOs could be used to demonstrate Bell-nonlocality \cite{TanWalColPRL91,HarPRL94}.
Recently, these  tests of Bell-nonlocality have been realized  \cite{HesEtalPRL04},   
as well as a different test using strong LOs \cite{BabEtalPRL04}. However, all of these experiments
relied on post-selection, which can be justified on the basis of the fair-sampling assumption for
inefficient photodetection in the experiments using weak LOs \cite{HesEtalPRL04}. 

The advantage of using a strong LO for homodyne measurement is that the high-intensity
detectors (photo-receivers) have an efficiency close to unity, as opposed to photon counters
which typically have a much lower efficiency. In Ref.~\cite{BabBreLvoPRL04}, the authors used
homodyne detection on a split single photon to demonstrate ``remote state preparation'',
which, they say, is a concept that ``can be traced back to the
seminal work of Einstein, Podolsky, and Rosen (EPR) \cite{EinEtalPR35},
who have considered an entangled state of two particles
with correlated positions and momenta. By choosing to
measure either the position or the momentum of her
particle, Alice can remotely prepare Bob's particle in
an eigenstate of either observable, thus instantaneously
creating either of two mutually incompatible physical
realities at a remote location.''

In fact, the EPR paper also considered a general
pure bipartite entangled state, and a general measurement by one party (Alice).
Moreover, in the same year, \sch\ generalized the EPR phenomenon to more
than two different measurement settings by Alice, and dubbed it ``steering'' \cite{SchPCP35}.
An experimentally testable criterion for the original (two-setting) EPR phenomenon
was developed by Reid \cite{ReiPRA89}. However, it was only in 2007 that a completely
general characterization of EPR-steering, for arbitrarily many measurements of arbitrary
type on an arbitrary bipartite state,  was developed by us and Doherty \cite{WisEtalPRL07}.
Even more recently, we and Cavalcanti and Reid \cite{CavEtalPRA09} have derived some
broad classes of experimental tests for EPR-steering.

EPR-steering is strictly easier to demonstrate than Bell-nonlocality \cite{WisEtalPRL07},
which has recently been shown experimentally using two-photon entangled states \cite{SauEtalNat10}. Thus one might
expect that the experiment in Ref.~\cite{BabBreLvoPRL04} did demonstrate this
effect for a single photon, {\em without post-selection} (unlike the above experiment \cite{SauEtalNat10}, which did use post-selection). The modern theory
discussed in the preceding paragraph provides the tools that
allow us, in this paper, to address this prospect.
We show that, unfortunately, the experimental imperfections of \cite{BabBreLvoPRL04} were too
great to allow a rigorous demonstration of EPR-steering
with a single photon. However, we also show that, with modest improvements in the experimental
parameters, and the addition of photon counting to the arsenal of Alice's measurements,
it should be possible to perform such a demonstration.

The remainder of the paper is organized as follows. First we model the experiment in Sec.~\ref{Model} and then we elaborate on the concept
of EPR-steering in Sec.~\ref{EPRSteering}.  We derive appropriate EPR-steering inequalities in Sec.~\ref{SteerInequalities} and establish
the conditions both sufficient and necessary to experimentally demonstrate EPR-steering in Sections \ref{SteerSufficient} and \ref{NecCond}
respectively.  In Sec.~\ref{Comparison} we compare our results with the experiments of Lvovsky and co-workers and conclude in Sec.~\ref{Summary}
with a summary of our findings.

\section{Modelling the Experiment}\label{Model}

\subsection{The single-photon state} \label{introstate}

Consider the case where Alice and Bob share
an entangled state formed from a single photon incident on a beam
splitter,  \beq
\ket{\psi^\chi}=\sqrt{\chi}\ket{0,1}-\sqrt{1-\chi}\ket{1,0},
\label{unevenstate} \eeq where $\chi$ is a variable defining the
beam splitter. For the special case of $\chi=0.5$,
\erf{unevenstate} is a maximally entangled state. However, in
realistic experiments the preparation of the initial photon is
probabilistic, occurring with efficiency $\eta$.
This means that in practice Alice and Bob will end up
with a mixed state that has a vacuum component.  That is, the
state that is actually prepared in such a situation has the form
\beq
W_\eta^\chi=\uplus\sqrt{1-\eta}\ket{0,0}\uplus\sqrt{\eta}\left(\sqrt{\chi}\ket{0,1}-\sqrt{1-\chi}\ket{1,0}
\right).\label{SingleUneven} \eeq
Here $\uplus$ is defined \cite{JonEtalPRA06} by the equation
$\uplus \alpha\ket{a} \equiv +|\alpha|^2\ket{a}\bra{a}$.

It is this type of state that was used by Lvovsky and co-workers
to demonstrate ``remote state preparation'' i.e. the EPR-steering phenomenon
\cite{BabBreLvoPRL04}, and to violate a Bell-inequality using post-selected
measurement results \cite{BabEtalPRL04}. The latter (more recent) experiment
had the better experimental parameters:  preparation of
states of the form of $W_\eta^\chi$ with $\eta=0.64$.

Since $W_\eta^\chi$ is a two-qubit entangled state, it is quite
straightforward to show that despite the introduction of the
vacuum component, the  state $W_\eta^\chi$ always retains at least
some of its entanglement for any nonzero $\eta$ (provided that
$\chi\neq 0,1$). In fact, one finds that this state possesses entanglement (concurrence) of
$E=2\eta\sqrt{\chi(1-\chi)}$ 
\cite{WooPRL98} (which simplifies to $E=\eta$ for $\chi=0.5$).

Clearly, for small $\eta$, the state
possesses little entanglement which may limit the usefulness of
$W_\eta^\chi$  for some QIP tasks. For instance, we can ask whether $W_\eta^\chi$
can be used to violate a Bell inequality.   For two-qubit states, there is
 an analytical test \cite{HorEtalPLA95}  for determining whether the state
violates the Clauser, Horne, Shimony, Holt (CHSH) inequality, the simplest sort
of Bell-inequality (and the sort tested in Ref.~\cite{BabEtalPRL04}).
This test reveals that it is necessary to have $\eta >
1/[2\sqrt{2\chi(1-\chi)}]$ to violate a CHSH-inequality.

As one might
expect, this is most easily satisfied when the initial
state possesses maximum entanglement, at $\chi=0.5$,
also as used in Ref.~\cite{BabEtalPRL04}.
This gives a necessary condition of $\eta > 1/\sqrt{2}$, compared to
the $\eta=0.64$ achieved in the experiment. This shows that CHSH-violation without
post-selection would have been impossible in this experiment. It is important
to note that $\eta > 1/\sqrt{2}$ is only a necessary condition --- even if it were
achieved in the experiment this does not mean that Bell-nonlocality could have
been demonstrated using the experimental detection techniques. First, the homodyne
detection did not have unit efficiency, and second it does not correspond to projective
measurements as are most useful for violating a CHSH-inequality. We turn in the
following subsection to describing the experimental detection scheme.

\subsection{Homodyne detection} \label{introhomo}

As discussed above, the experiments \cite{BabBreLvoPRL04,BabEtalPRL04} use the
high-efficiency measurement technique of homodyne detection with a strong LO.
EPR-steering is about whether Alice's measurements affect Bob's
state (in a sense to be defined rigorously later); the only efficiency that matters
is Alice's.  Specifically, all we need to know is Bob's state conditioned on
 Alice's measurement results. We can allow for the non-unit efficiency $\eta_{\rm h}$
 of Alice's
 measurements by introducing  a finite probably $1-\eta_{\rm h}$ of photon
loss at Alice's side prior to her measurement.  Thus, we
can modify the state $W_\eta^\chi$ to include this loss, then
proceed using the measurement formalism for perfect efficiency homodyne detection.

We can describe loss by the two Kraus operators, corresponding to losing and not losing a photon 
respectively, 
\bqa
\hat{M}_{\rm lose}&=&\sqrt{1-\eta_{\rm
h}}\ket{0}_A\bra{1},\label{Mloss}\\
\hat{M}_{\rm keep}&=&\sqrt{\hat{1}-\hat{M}^\dagger_{\rm lose}\hat{M}_{\rm
lose}}\nonumber \\
&=&\ket{0}_A\bra{0}+\sqrt{\eta{\rm
_h}}\ket{1}_A\bra{1},\label{Mnoloss}
\eqa
where the $A$ subscript reminds us that this is for Alice's mode.
Therefore, the effective state allowing for Alice's inefficient detection is
\bqa
 W^{\chi \rm hom}_{\eta}&=&\hat{M}_{\rm lose}W_\eta^\chi\hat{M}^\dagger_{\rm lose}+\hat{M}_{\rm keep}W_\eta^\chi\hat{M}^\dagger_{\rm keep}  \nonumber\\ 
&=&[(1-\eta)+(1-\eta_h)\eta(1-\chi)]\ket{00}\bra{00}\nonumber\\&&\phantom{ =}\!\!\!\!\!\!+\eta\eta_{\rm h}(1-\chi)\ket{10}\bra{10}+\eta\chi\ket{01}\bra{01}\nonumber\\
&&\phantom{ =}\!\!\!\!\!\!-\sqrt{\eta_{\rm
h}}\eta\sqrt{\chi(1-\chi)}\left(\ket{10}\bra{01}+\ket{01}\bra{10}\right).\nonumber\\
\eqa 

 The effect operator for homodyne measurement using a LO of phase $\theta$
on a state with at most one photon is
\cite{WisQSO95}
\beq\hat{F}^\theta_{\rm
hom}(r)=\frac{\exp(-r^2/2)}{\sqrt{2\pi}}\left( \ket{0}\bra{0} +r
\hat{\sigma}_\theta + r^2\ket{1}\bra{1}\right).\label{Fr} \eeq
Here $\hat{\sigma}_\theta$ is defined as in
\beq
\hat{\sigma}_\theta=\cos(\theta)\hat{\sigma}_x+\sin(\theta)\hat{\sigma}_y ,
\eeq
where the Pauli operators are defined in the usual way given $\s{z} = \ket{1}\bra{1}-\ket{0}\bra{0}$,
where here $\ket{0}$ and $\ket{1}$ are Fock states. These operators define a POVM normalized as
$\int_{-\infty}^\infty dr \hat F(r) = \hat{1}$, and the measurement result $r$ is the suitably integrated homodyne photocurrent \cite{WisQSO95,WisKil97}. Thus if Alice makes a homodyne measurement with result $r$,
 Bob's conditioned state is
\beq
\tilde{\rho}_B^{\theta}(r)={\rm Tr}_A[\hat{F}^\theta_{\rm hom}(r)
W_\eta^{\chi \rm hom}]
\label{bobsstate}
\eeq
Here the tilde denotes an unnormalized state, the norm of which equals the probability density for Alice
to obtain the result $r$.


\subsection{Photodetection} \label{introphoto}

Although the experiments \cite{BabBreLvoPRL04,BabEtalPRL04} used only homodyne detection,
we will see later that, for the purposes of EPR-steering, it can be useful to also
consider photon counting, even though that typically has a far lower efficiency, $\eta_{\rm p}$.
In this case loss is very easy to include in the description of the measurement itself,
which is  described by the photodetection effect operators 
\bqa
\hat{F}_+ &=&\eta_{\rm p}\ket{1}\bra{1}, \\
\hat{F}_-&=&\hat{1}-\hat{F}_+.
\eqa 
corresponding to detecting and not detecting a photon respectively.
This time Bob's conditioned states are 
\bqa
\tilde{\rho}_B^{\rm p}(+)&=&{\rm Tr}_A[\hat{F}_+W_\eta^{\chi}]  = \wp_+ \frac{1}{2}(\hat 1 + z_+\s{z}) ,\\
\tilde{\rho}_B^{\rm p}(-)&=&{\rm Tr}_A[\hat{F}_-W_\eta^{\chi}]  = \wp_- \frac{1}{2}(\hat 1 + z_-\s{z}).
\eqa
That is, the states are mixtures of $\s{z}$ eigenstates (Fock states), with
\bqa
\wp_+&=& \eta\eta_{\rm p}(1-\chi),\nonumber\\
\wp_-&=& 1-\eta\eta_{\rm p}(1-\chi),\nonumber\\
z_+&=&\frac{\langle \hat{F}_+\otimes \hat{\sigma}_z\rangle}{\wp_+}= -1, \nonumber\\
z_-&=&\frac{\langle \hat{F}_-\otimes
\hat{\sigma}_z\rangle}{\wp_-}=\frac{2\eta\chi-(1-\eta\eta_{\rm p}(1-\chi))}{1-\eta\eta_{\rm p}(1-\chi)}.\nonumber\\
  \eqa

\section{EPR-Steering}\label{EPRSteering}

\subsection{Defining EPR-Steering}

The concept of steering introduced by \sch\ in 1935
\cite{SchPCP35} as a generalization of the Einstein-Podolsky-Rosen
(EPR) paradox has received renewed interest in recent years
(see for example Refs.~\cite{KirFPL06, WisEtalPRL07, JonWisDohPRA07, CavEtalPRA09,ReiEtalRMP09,SauEtalNat10,OppSCI10}). 
In particular, it was given a formal definition
\cite{WisEtalPRL07} as a quantum information task involving two parties,  Alice and Bob. They
share a bipartite quantum state, and Alice's task is to convince Bob
that it is entangled (assuming that it is) even though Bob does not trust her.
Alice can try to convince Bob that the state is entangled if she can
`steer' Bob's system into different ensembles of states by making different
measurements on her part of the state, by virtue of the
entanglement and the EPR effect. Bob will only be convinced however if
the results he obtains could not be described by a local
hidden state (LHS) model. That is, he must rule out the possibility that Alice is simply
sending him a pure state, drawn from some ensemble, and using her knowledge of his state
to pretend to be able to steer it. Thus we define the experiment to be a demonstration of
the EPR-steering phenomenon if and only if (iff) Bob is convinced that the state
is entangled.

We can make the above definition more formal as follows. To connect more directly with the rest of this
paper (and with experiment) we give a slightly less general formulation than that in Ref.~\cite{WisEtalPRL07}. 
Alice and Bob make measurements
on their subsystems. Because Bob trusts his own devices, there is no necessity for
his measurement to be efficient. In fact,  we do not describe his measurement process explicitly (this is the point of difference from Ref.~\cite{WisEtalPRL07}), but
simply assume that he is able to make measurements that enable him to determine the average of
some set $\scu{\hat{B}_j}$ of observables (acting on his subsystem alone) from an ensemble of repeated experiments.
In a given run, Bob decides which $\hat{B}_j$ he is interested in, and {\em after receiving his subsystem},
informs Alice of his choice. Alice then makes a measurement on her subsystem which we can describe without loss of
generality by some observable $\hat{A}_j$ (in the case of generalized measurements, this operator would have
to be considered to act on an ancilla  as well as her subsystem). We denote Alice's outcome by $a_j$, a random variable
taking the eigenvalues of $\hat{A}_j$ as its possible values. Bob can then calculate the average $\san{\hat B_j}_{a_j}$ of
$\hat{B}_j$ from each subensemble corresponding to the different outcomes $a_j$ of Alice.

Now if there is a LHS ensemble for Bob,
described by Bob-states $\rho_\xi$ with weights $\wp(\xi)$, it must be the case that, for all $j$,
\beq \label{formofBA}
\san{\hat B_j}_{a_j} = \sum_\xi \wp(\xi|a_j) \tr{\rho_\xi \hat B_j}
\eeq
where $\wp(\xi|a_j) = \wp(\xi)\wp(a_j|\xi) / \sum_\xi \wp(\xi)\wp(a_j|\xi) $. Here
$\wp(a_j|\xi)$ is the probability that Alice (here assumed by Bob to be trying to cheat) announces
the result $a_j$ when she knows Bob's state to $\rho_\xi$. Thus if Bob's set of expectation values
$\scu{\san{\hat B_j}_{a_j}}$, for all $j$ and all $a_j$, are {\em not} consistent with the form
of \erf{formofBA} then he has to admit that Alice cannot be cheating. That is, she has demonstrated
EPR-steering of his state, and the state they share must be entangled.

It was shown in \cite{WisEtalPRL07} that there exist states that cannot violate any Bell inequality,
but which do allow EPR-steering to be demonstrated. In particular, this was the case
for two-qubit Werner states with a mixing parameter  between  0.5 and $0.66$.
This  gives hope  that the mixed states $W^\chi_\eta$ of interest in this paper could also be steerable
even with the experimental $\eta = 0.64$. To be useful experimentally, however, what we require
is an inequality involving measurable quantities, analogous to a Bell inequality,
that, if violated, would demonstrate the EPR-steering phenomenon.
The first inequality of this nature was introduced by Reid \cite{ReiPRA89}. A rigorous
derivation of a number of broad classes of these {\em EPR-steering} inequalities was first given in Ref.~\cite{CavEtalPRA09}.
In the following subsection we review the class of inequality we require
for this paper.

If a quantum state violates an EPR-steering inequality
that means that it cannot be that there exists an ensemble of local hidden states (LHSs) for Bob's
subsystem that will explain the observed correlations.  That is,
violation of an EPR-steering inequality is a sufficient condition for
demonstrating EPR-steering.  In many cases we can find explicitly a LHS model which
saturates the bound of the inequality. In such cases we will say that the the EPR-steering
inequality is {\em tight}.

\subsection{Additive Convex EPR-steering inequalities} \label{AddConvex}

A general approach to deriving EPR-steering inequalities is to begin
with a constraint which holds for Bob's system given that it is
described by a quantum state.   A particularly
useful type of constraint on Bob's system for determining EPR-steering
inequalities are those constraint which take an additive, convex
form.   The convexity of the constraints on Bob's
system is the key feature which allows derivation of the
inequalities.

Consider the case where Bob's expectation values are constrained
(by the assumption that they are derived from a quantum system) by
an inequality of the following form:
\beq \sum_j f_j(\langle \hat{B}_j\rangle,\alpha_j) \leq
c, \ \ \ \ \ \forall \alpha_j \in \lambda
(\hat{A}_j),\label{constraint} \eeq
where $f_j$ is a convex
function of the variable $\langle \hat{B}_j\rangle$ and
$\lambda(\hat{A}_j)$ are the eigenvalues of operator $\hat{A}_j$.
We term this an \emph{additive convex} constraint.  The convexity
property of $f_j$ means that inequalities of the
following type must be satisfied, \beq f_j(px+(1-p)y,\alpha)\leq
pf_j(x,\alpha)+(1-p)f_j(y,\alpha), \label{star} \eeq
for all $p \in [0,1]$.

Now if Bob's system possesses a LHS, then  we
know that  Bob's expectation
value given Alice's result $a_j$ is given by \erf{formofBA}, which we rewrite as
\beq
\langle  \hat{B}_j\rangle_{a_j}=\sum_\xi \wp(\xi|a_j)\langle
\hat{B}_j\rangle_{\rho_\xi}.\label{define} \eeq
Using this definition for $\langle  \hat{B}_j\rangle_{a_j}$ and
recalling the convexity property of $f_j$ one finds
\beq
f_j(\langle \hat{B}_j\rangle_{a_j},\alpha_j) \leq
\sum_{\xi}\wp(\xi|a_j)f_j(\langle
\hat{B}_j\rangle_{\rho_\xi},\alpha_j).\label{sumineq} \eeq

Now consider the following expectation value involving Bob's expectation
values and Alice's measurement results:
\beq E_{A_j}[f_j(\langle
\hat{B}_j\rangle_{a_j},a_j)]=\sum_{a_j}\wp(a_j)f_j(\langle
\hat{B}_j\rangle_{a_j},a_j).\label{expect1} \eeq Substituting
for $\langle  \hat{B}_j\rangle_{a_j}$ and using \erf{sumineq} we
find \beq E_{A_j}[f_j(\langle \hat{B}_j\rangle_{a_j},a_j)] \leq
\sum_{\xi,a_j}\wp(\xi,a_j)f_j(\langle
\hat{B}_j\rangle_{\rho_\xi},a_j). \label{expect2} \eeq
Taking a sum over the possible measurements $j$ \beq \sum_j
E_{A_j}[f_j(\langle \hat{B}_j\rangle_{a_j},a_j)] \leq
\sum_{\xi,a_j}\wp(\xi,a_j)\sum_j f_j(\langle
\hat{B}_j\rangle_{\rho_\xi},a_j) \label{a1}, \eeq and finally,
using the initial constraint \erf{constraint} gives \beq \sum_j
E_{A_j}[f_j(\langle \hat{B}_j\rangle_{a_j},a_j)] \leq c \label{a}.
\eeq

Thus we have arrived at an EPR-steering inequality, the {\em violation}
of which is an experimental criterion for \emph{demonstrating}
EPR-steering.  This is a condition which allows detection of EPR-steering
of Bob's state based on measured expectation values for his system
(conditioned on the results Alice reports), and the results Alice reports.
Note that in deriving this inequality no assumption was made that Alice's
results derived from the measurement of a quantum system; this is necessary
for a skeptical Bob to be convinced.

\section{EPR-steering inequalities for a qubit}\label{SteerInequalities}

In this section we apply the general formalism of the preceding section
to derive EPR-steering inequalities for a qubit, as describes Bob's half
of a split single photon.

\subsection{Linear inequality for an infinite number of measurements}\label{SingleInfiniteCriteria}

A special case of additive convex EPR-steering inequaities is that of \emph{linear}
inequalities. Here we consider such inequalities for an infinite number of different
observables by Bob, which we call equatorial observables.
By this we mean observables defined by axes around the $z=0$ plane of the Bloch sphere.
Hence we consider the following operator sum for
Bob's system \beq \hat{S}_{\rm
plane}=\frac{1}{\pi}\int_{-\pi/2}^{\pi/2}d\theta
\alpha_\theta\hat{\sigma}_\theta, \label{BobOperator}\eeq where
$\hat{\sigma}_\theta=\cos(\theta)\hat{\sigma}_x+\sin(\theta)\hat{\sigma}_y$ as before.
We only consider the half-plane $|\theta|\leq \pi/2$ because $\s{\theta} = - \s{\theta+\pi}$
and so these are not distinct observables.
 We assume that  for all $\theta$, $\alpha_\theta \in \{-1,1\}$, the same set of 
 possible values as the 
measurement outcomes $a_\theta\in \{-1,1\}$  for the measurement $\hat{A}_\theta$ that Alice
performs on being informed of Bob's choice of $\theta$. Note that while Alice must (or at least,
should, in her own best interest) make a different measurement for each $\theta$, Bob can determine
the average of $\s{\theta}$ for any $\theta$ by sometimes measuring $\s{x}$
and sometimes $\s{y}$ (or, with more relevance to the split-single-photon case, by making
any set of tomographically complete measurements, such as homodyne measurements \cite{BabEtalPRL04}).

For any quantum state for Bob's system, the expectation value
of the operator $\hat{S}_{\rm plane}$ will be bounded: \beq
\langle \hat{S}_{\rm plane}\rangle =
\frac{1}{\pi}\int_{-\pi/2}^{\pi/2}d\theta \alpha_\theta\langle
\hat{\sigma}_\theta\rangle \leq c_{\rm plane}, \eeq
  where the bound is obtained by calculating
\beq c_{\rm plane} = {}^{\rm max}_{\{\alpha_\theta\}}\lambda_{\rm
max}(\hat{S}_{\rm plane}). \eeq
Clearly the maximum
over $\{\alpha_\theta\}$ occurs when $\alpha_\theta =1 \, \forall \theta$.
Under this condition, we simply need to perform the integral and calculate
the maximum eigenvalue, which results in $c_{\rm plane}={2/\pi}$
and thus \beq \langle \hat{S}_{\rm plane}\rangle =
\frac{1}{\pi}\int_{-\pi/2}^{\pi/2}d\theta \alpha_\theta\langle
\hat{\sigma}_\theta\rangle\leq
\frac{2}{\pi}.\label{PlaneConstraint} \eeq

Due to the additivity of the integral operation and the convexity
of the expectation value we have arrived at a constraint on Bob's
system which takes an additive, convex form. Thus, using the
method of Sec.~\ref{AddConvex}, one can derive from
\erf{PlaneConstraint} the following EPR-steering inequality \beq
\frac{1}{\pi}\int_{-\pi/2}^{\pi/2}d\theta
E_{A_\theta}[a_\theta\langle
\hat{\sigma}_\theta\rangle_{a_\theta}] \leq
\frac{2}{\pi}.\label{PlaneIneq1} \eeq

Noting that  the conditional expectation
value $E_{A_\theta}[a_\theta\langle
\hat{\sigma}_\theta\rangle_{a_\theta}]$ can be more simply expressed as $\langle
\hat{A_\theta}\hat{\sigma}_\theta\rangle$, we can rewrite
\erf{PlaneIneq1} as \beq
\frac{1}{\pi}\int_{-\pi/2}^{\pi/2}d\theta\langle
\hat{A_\theta}\hat{\sigma}_\theta\rangle \leq \frac{2}{\pi}.
\label{SingleInequality2} \eeq
The violation of this inequality by the
measured correlations on a bipartite
quantum state would be a demonstration of EPR-steering.
We will address the problem that it is not really possible for Alice
to perform an infinite number
of different measurements in an experiment in Sec.~\ref{finite_sec}.

\begin{figure}[]
\begin{center}
\includegraphics[width=8.5cm]{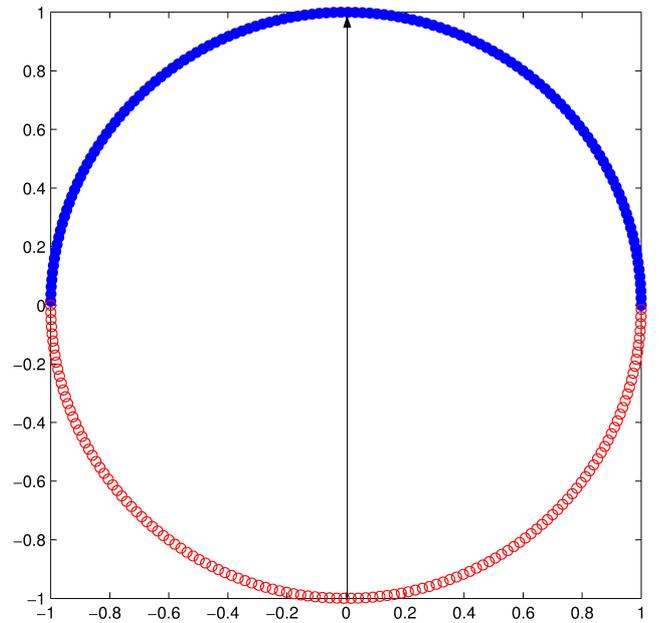}
\end{center}
\caption[Partitioning an equatorial ensemble]{[Color online] An infinite LHS
ensemble on the equator of the Bloch sphere.  If the measurement axis is in
the direction of the black vector, then Alice would report results $-1$ or $+1$
depending on whether the state she has sent in the  aligned (blue solid circles) or antialigned (red circles) subensemble.}
\label{CircleEnsemble}
\end{figure}

We now exhibit a simple LHS model to simulate correlations of the form
$\langle \hat{A_\theta}\hat{\sigma}_\theta\rangle$ around the equator
of the Bloch sphere.  Due to the symmetry of the measurement
arrangement, a suitable ensemble would consist of an infinite
number of pure states $\ket{\xi}$ on the $z=0$ unit circle.
For any  measurement axis $\theta$, the
ensemble can be partitioned into two even halves as shown in
Fig.~\ref{CircleEnsemble}.  When Bob reveals his axis $\theta$,
then, Alice could report the $+1$ result if she sent a
pure state closer to the positive measurement axis, or $-1$ result
for a state closer to the negative axis. It is easy to verify that  such a scheme would
give the correlation $\langle \hat{A_\theta}\hat{\sigma}_\theta\rangle=2/\pi$.
Thus the inequality (\ref{SingleInequality2}) is {\em tight}.

\subsection{Nonlinear EPR-steering
inequality}\label{sec:nonlinearSI}

As discussed in Ref.~\cite{CavEtalPRA09},
nonlinear EPR-steering inequalities are in general better able to detect experimental
steerability than simple linear inequalities.  As we will show, that is the case here.
In order to derive
the inequality we make use of the linear inequality of the
previous section involving equatorial observables,
and augment it with a single additional non-equatorial Bob-observable, $\s{z}$.
Once again, this involves no more extra work on Bob's behalf, if he is already making
a set of tomographically complete measurements, such as homodyne measurements \cite{BabEtalPRL04}.
For Alice, it is in her best interests now to make a different measurement, $\hat{A}_z$,
whenever Bob tells her that in this run he is interested in $\s{z}$. For simplicity we assume that
$\hat{A}_z$ also has two possible outcomes: $a_z\in \{-1,1\}$. We show below,  however,
that even if Alice makes no measurement in this case, and merely always
reports $a_z=+1$  (for instance),  the inequality we derive is still stronger in general than the preceding one,
\erf{SingleInequality2}.

Consider the following function
\beq f(\{
\langle \hat{\sigma}_\theta \rangle\},\langle
\hat{\sigma}_z\rangle)=\frac{1}{\pi}\int_{-\pi/2}^{\pi/2}d\theta
a_\theta\langle \hat{\sigma}_\theta\rangle -
\frac{2}{\pi}\sqrt{1-\langle
\hat{\sigma}_z\rangle^2},\label{fconvex} \eeq  which, as shown in
Appendix~\ref{app}, is a convex function of its arguments and
satisfies $f \leq 0\ \forall\ a_\theta\in\{-1,1\}, \forall\ \rho$.
Therefore, the constraint $f(\{ \langle \hat{\sigma}_\theta
\rangle\},\langle \hat{\sigma}_z\rangle)\leq 0$ defines an
additive convex constraint on Bob's system and using the approach
of Sec.~\ref{AddConvex} leads to the nonlinear EPR-steering inequality
\beq \frac{1}{\pi}\int_{-\pi/2}^{\pi/2}d\theta
E_{A_\theta}[a_\theta\langle
\hat{\sigma}_\theta\rangle_{a_\theta}]-\frac{2}{\pi}E_{A_z}\sq{\sqrt{1-(\langle
\hat{\sigma}_z\rangle_{a_z})^2}} \leq 0.\label{NLIneq} \eeq
Here $A_z$ ($A_\theta$) denotes the measurement Alice performs (or, as far as Bob is concerned,
 purports to perform) when Bob reveals that he has measured $\s{z}$ ($\s{\theta}$).
Noting
that the first term is the same as \erf{PlaneIneq1},  and rearranging the inequality gives \beq
\frac{1}{\pi}\int_{-\pi/2}^{\pi/2}d\theta\langle
\hat{A_\theta}\hat{\sigma}_\theta\rangle\leq
\frac{2}{\pi}E_{A_z}\sq{\sqrt{1-\langle \hat{\sigma}_z\rangle_{a_z}^2}}.
\eeq

Finally, using our  assumption that Alice's observable $\hat{A}_z$ is
dichotomic, we can write out the conditional expectation on the
right hand side explicitly to obtain the nonlinear EPR-steering
inequality \beq \frac{1}{\pi}\int_{-\pi/2}^{\pi/2}d\theta\langle
\hat{A_\theta}\hat{\sigma}_\theta\rangle \leq
\frac{2}{\pi}\left[\wp_+\sqrt{1-z_+^2}+
\wp_-\sqrt{1-z_-^2}\right], \label{NLInequality} \eeq
where
$\wp_\pm$ is the probability that Alice
obtains results $\pm 1$ and $z_\pm=\langle {\s{z}} \rangle_\pm$
are Bob's respective conditional expectation values.

In general, the nonlinear bracketed term on the right hand side of
\erf{NLInequality} will be less than 1 and hence this side of the
inequality will be less than $2/\pi$.  Thus, as expected,
the nonlinear EPR-steering inequality which incorporates
an additional measurement setting, will be generally easier to
violate than \erf{SingleInequality2}. Note that even if Alice's observable
is trivial ($\hat{A}_z = \hat 1$), so that there is only one ``result'' ($\wp_+=1$)
we obtain
\beq \frac{1}{\pi}\int_{-\pi/2}^{\pi/2}d\theta\langle
\hat{A_\theta}\hat{\sigma}_\theta\rangle \leq
\frac{2}{\pi} \sqrt{1-\tr{\s{z}\rho}^2}, \label{trivialNLInequality} \eeq
where $\rho$ is Bob's unconditioned reduced state.
The right hand side here is still less than that of \erf{SingleInequality2}, except in the
case that $\tr{\s{z}\rho} = 0$. This is the only case where \erf{NLInequality}
is not stronger than \erf{SingleInequality2}.

This example is similar to the so-called `inept state' example of
Ref.~\cite{JonEtalPRA05} which suggests the following LHS model to model the correlations
considered here. Consider two rings of pure
states on the Bloch sphere centered around the $z$ axis, at $z=z_-$
and $z=z_+$.  The ensemble is
weighted so that it is invariant under rotations around the $z$-axis, and that
so that the weighting of the upper (lower) ring is given by $\wp_+$
($\wp_-$). By construction, such an ensemble will produce the
correct correlations when Bob measures
$\hat{\sigma}_z$, if Alice announces $a_z=+1$ ($a_z=-1$) when the LHS state
she has sent is in the upper (lower) ring.
  Thus, we simply need to determine how well such
an ensemble simulates correlations of the form  $\langle
\hat{A_\theta}\hat{\sigma}_\theta\rangle$.  A straightforward
calculation shows that the radii of the rings of pure states are
given by $(1-z_\pm^2)^{1/2}$.  Taking the
average over half of each ring centered around the
measurement axis of interest results in the integral of
\erf{SingleInequality2}, multiplied by each ring's radius.  Taking the
weighted average, this LHS model can simulate a value for the left hand side of
\erf{NLInequality} equal to
\beq
\frac{2}{\pi}\left(\wp_+\sqrt{1-z_+^2}+
\wp_-\sqrt{1-z_-^2}\right).
\eeq
This proves that this ensemble of
LHSs is optimal for these observables, and hence that \erf{NLInequality}
is a tight inequality.

\subsection{Finite setting inequality}\label{finite_sec}
The inequalities derived in the previous subsections assume that Alice
can make  an infinite number of measurements: a different
$\hat{A}_\theta$ for each value of $\theta$ tested  by Bob.
In practice, a realistic experiment will be constrained to some
finite number of measurement settings, such as a finite set of $\theta$ values.
One might
expect that using a finite number of settings will make it more
difficult to demonstrate EPR-steering.  While this is indeed the case,
we show that we can modify \erf{NLInequality} to account for this,
and that the increase in difficulty is small even for moderate values of $n$.

Consider the case analogous to Sec.~\ref{SingleInfiniteCriteria}
 but with $n$ evenly spaced equatorial measurements,  which implies that
the $\{\theta_i\}$ are separated by $\pi/n$.  Assuming a LHS model
for Bob's subsystem ensures \beq \langle \hat{S}_{\rm
plane}^{(n)}\rangle = \frac{1}{n}\sum_{i=1}^{(n)} \alpha_i\langle
\hat{\sigma}_{\theta_i}\rangle \leq f(n),\label{finiteplane} \eeq
where the bound is a function of the number of measurement axes
and is given by \beq f(n)={}^{\rm max}_{\{\alpha_i\}}\lambda_{\rm
max}(\hat{S}_{\rm plane}^{(n)}). \eeq
Following the approach used in
the examples of Ref.~\cite{SauEtalNat10}, one finds that the
eigenvectors associated with the maximum eigenvalues obtainable
for $\hat{S}_{\rm plane}^{(n)}$ occur along the direction of the
measurement axes and a direction midway between measurement axes
for $n$ odd and even respectively. Calculating these eigenvalues
for the first few small $n$ allows one to obtain by induction \beq
f(n)=\frac{1}{n}\left(\left|\sin\left(\frac{n\pi}{2}\right)\right|+2\sum_{k=1}^{\lfloor
n/2\rfloor}\sin\left[\left(2k-1\right)\frac{\pi}{2n}
\right]\right)\label{finite}. \eeq Proceeding analogously to
Sec.~\ref{SingleInfiniteCriteria}, the constraint
\erf{finiteplane} results in the EPR-steering inequality \beq
\frac{1}{n}\sum_{i=1}^{ n} \langle
\hat{A}_i\hat{\sigma}_{\theta_i}\rangle \leq f(n).
\label{FiniteInequality1} \eeq

We can also incorporate an additional measurement axis
orthogonal to the $n$ equatorial measurements,
analogously to the proof in Sec.~\ref{sec:nonlinearSI}.
We find that \erf{NLInequality} is modified to
\beq
\frac{1}{n}\sum_{i=1}^n\langle
\hat{A}_i\hat{\sigma}_{\theta_i}\rangle \leq
f(n)\left[\wp_+\sqrt{1-z_+^2}+
\wp_-\sqrt{1-z_-^2}\right].\label{SingleInequality4} \eeq
Also following the method of Sec.~\ref{sec:nonlinearSI} it is easy
to see that the a two-ring LHS construction is optimal, but now with each ring containing a finite
number of evenly space pure states. It follows on average the finite ensembles will predict \beq
\frac{1}{n}\sum_{i=1}^n\langle
\hat{A}_i\hat{\sigma}_{\theta_i}\rangle =
g(n)\left[\wp_+\sqrt{1-z_+^2}+
\wp_-\sqrt{1-z_-^2}\right],\label{LHSfinite} \eeq
To prove that \erf{SingleInequality4} is tight it remains
for us to prove that the functions $g(n)$ are the
same as the $f(n)$ of \erf{finite} for the optimal arrangement
for the pure states in the LHS ensemble.

\begin{figure}[ht]
\begin{center}
\includegraphics[width=8.5cm]{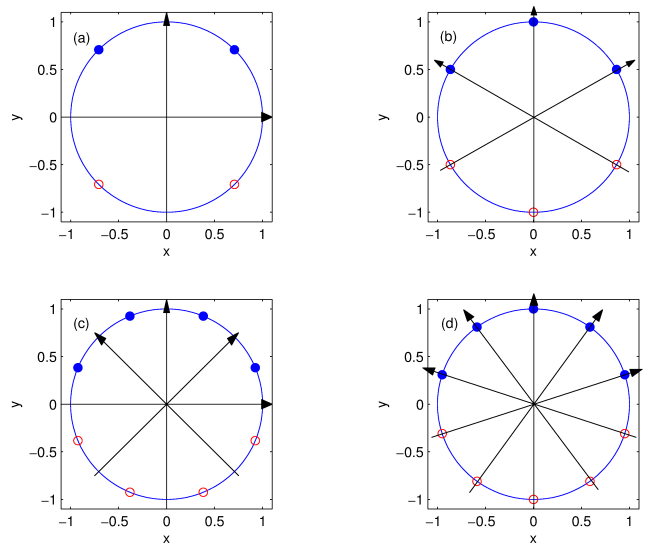}
\end{center}
\caption[Partitioning of finite ensembles]{[Color online]
Finite LHS ensembles
around a circular plane. In each case the black vectors denote $n$
measurement axes, and the coloured dots denote pure states.  Plots
(a) and (c) correspond to the first two even cases, $n=2$ and
$n=4$ respectively and have optimal ensembles of states lying
midway between the measurement axes.  Plots (b) and (d) correspond
to the first two odd cases, $n=3$ and $n=5$ respectively and have
optimal ensembles aligned with the measurements axes.  In each
case the blue (solid) and red colouring of the pure states denotes the
partitioning of the ensemble that Alice would use if the $y$-axis
was the measurement of interest for a particular run of an
experiment. } \label{finiterings}
\end{figure}

Consider the first few even and odd examples for $n$ as shown in
Fig.~\ref{finiterings} (in the diagram a single ring is
shown, but the optimal arrangement of LHSs will be the same for
both rings). For the first even cases, $n=2$ and $n=4$, the
ensembles that minimize $g(n)$ have pure states lying midway between the
measurement axes (this reflects the directions of the maximum
eigenvectors of $\hat{S}_{}^{(2)}$ and $\hat{S}_{}^{(4)}$).  For the
first odd cases, $n=3$ and $n=5$, the optimal  ensembles have
pure states aligned with the measurement axes (reflecting the
directions of the maximum eigenvectors of  $\hat{S}_{}^{(3)}$ and
$\hat{S}_{}^{(5)}$).  Partitioning each of these ensembles in half
(as indicated in Fig.~\ref{finiterings}) leads to
$g(2)=1/\sqrt{2}$, $g(3)=2/3$, $g(4)\approx 0.6533$, and
$g(5)\approx 0.6472$.  It is straightforward to show that these
results generalise as one would expect for larger $n$ (with
ensembles of $2n$ pure states off-axis per ring for even $n$, and
$2n$ pure states on-axis per ring for odd $n$). Moreover,  $g(n)=f(n)$ for all $n$. Thus,
\erf{LHSfinite} saturates the right hand side of
\erf{SingleInequality4} and so the latter is indeed a tight EPR-steering inequality.

\begin{figure}[]
\begin{center}
\includegraphics[width=8.5cm]{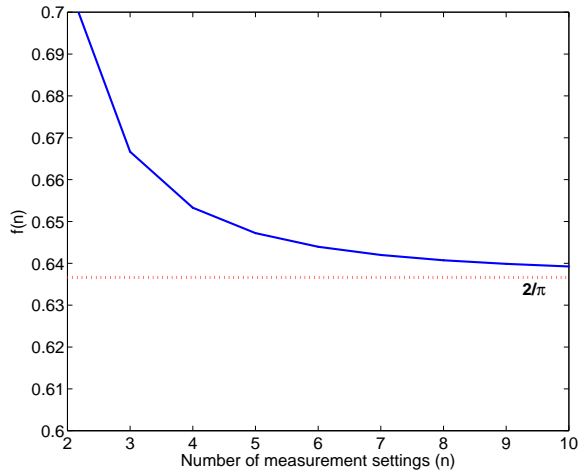}
\end{center}
\caption[Finite measurement correction factor]{The finite correction factor $f(n)$ for some small values of the number of equatorial measurements used, $n$.  We see that $f(n)$ quickly approaches the asymptotic value of $2/\pi$ (indicated by the red dotted line).}
\label{f(n)}\end{figure}

It can be seen in Fig.~\ref{f(n)} that the function $f(n)$ quickly
approaches its asymptotic value.  Hence, it takes relatively few
equatorial measurement settings to arrive at an EPR-steerability criterion
which works almost as well as the ideal criterion which
incorporates an infinite number of measurement settings.
Note that in the limit $n\rightarrow \infty$, $f(\infty) \to 2/\pi$ and
\erf{SingleInequality4} is equivalent to \erf{NLInequality}.

\section{EPR-steering of a single photon}\label{SteerSufficient}
\subsection{Evenly split single photon state}\label{Even}

We will now apply the EPR-steering inequalities derived in the preceding
section to the single-photon state introduced in Sec.~\ref{introstate},
with Alice's measurements constrained (as in experiment) to homodyne detection
as introduced in Sec.~\ref{introhomo}, supplemented by photon counting
as introduced in Sec.~\ref{introphoto}. We
begin by considering the case where $\chi=0.5$, that is,
when the intended initial state $W_\eta^\chi$ is maximally entangled
in the limit $\eta \to 1$.

Consider the most general case using the nonlinear EPR-steering
inequality defined in \erf{SingleInequality4}, as this allows for
both homodyne ($A_\theta$) and photodetection ($A_z$)
measurements.  We must take into
account the homodyne measurement inefficiency when evaluating the
left hand side of the inequality; the photodetection inefficiency
will manifest in the right hand side of the inequality.

Thus we wish to
calculate the maximum value of
$\frac{1}{\pi}\int_{0}^{\pi}d\theta\langle
\hat{A_\theta}\hat{\sigma}_\theta\rangle$ allowed by quantum
theory, given that Alice is restricted to homodyne measurements.
However, this may be simplified by noting that in the state $W_\eta^\chi$, for every
angle $\theta$ the maximum value of the correlation will be
obtained when Alice chooses the phase of the local oscillator to be $\theta$,
and reports a result $a_\theta=a(r)$, where $a(r)$ is negatively related to her homodyne
measurement result $r$.
This will result in the same value for the correlation
function for every direction.  Hence the correlation function
becomes independent of $\theta$ and we have
\beq
 \frac{1}{\pi}\int_{-\pi/2}^{\pi/2}d\theta\langle
\hat{A}_\theta\hat{\sigma}_\theta\rangle =\max_{a(r)}\left(\int dr{\rm Tr}[\tilde{\rho}_B^0(r)\hat{\sigma}_x a(r)] \right),\label{maxcorr}
\eeq where $\tilde{\rho}_B^0(r)$ is Bob's conditioned state defined
in \erf{bobsstate}, evaluated for $\theta = 0$,  and the function $a(r)$ is Alice's
reported result.  Recall that in deriving the ineqality (\ref{SingleInequality4}) we assumed
that $a(r) \in \{-1,1\}$, but apart from that restriction,  Alice is free to report any function of her
result $r$.  Using \erf{bobsstate},  \erf{maxcorr} evaluates
to
\beq \max_{a(r)}\left(-\eta\sqrt{\eta_{\rm h}}\int_{-\infty}^\infty
dr \frac{r}{\sqrt{2\pi}}\exp(-r^2/2)a(r) \right). \eeq
The maximum over $a(r)$ occurs
when Alice chooses $a(r)=-{\rm sign}(r)$ and thus we have
\bqa
\frac{1}{\pi}\int_{0}^{\pi}d\theta\langle \hat{A_\theta}\hat{\sigma}_\theta\rangle 
&=&\sqrt{\frac{2}{\pi}}\,\eta\sqrt{\eta_{\rm
h}}.\label{QMSinglesteer} \eqa
Recall that $\eta$ is the efficiency of production of the single photon, while $\eta_{\rm h}$
is the efficiency of Alice's homodyne measurement.  It is easy to verify that exactly the same equation holds when the left-hand-side 
of \erf{QMSinglesteer} is replaced by a finite sum, as in the left-hand-side of \erf{SingleInequality4}. 

\begin{figure}[]
\begin{center}
\includegraphics[width=8.5cm]{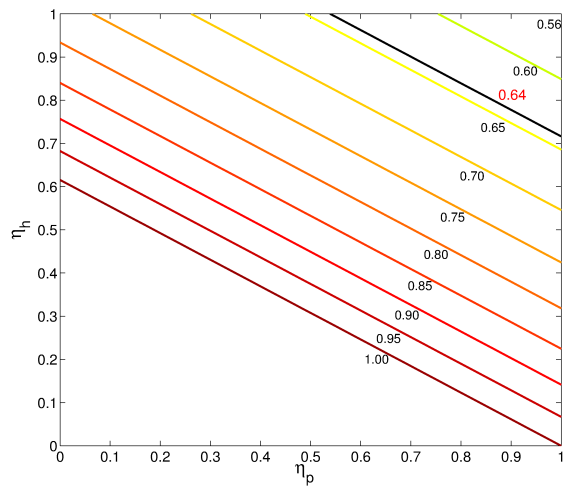}
\end{center}
\caption[Steering bound with inefficient homodyne detection]{The
pale contours indicate the value of $\eta$ required to demonstrate
steering with inefficient homodyne and photodetection
measurements.  For lower measurement efficiencies there are no
contours, as it would require unphysical values of $\eta$ (i.e.
$\eta > 1$) to satisfy the steerability criterion.  The black
contour ($\eta=0.64$) marks the preparation efficiency limit
achieved experimentally in \cite{BabEtalPRL04}.}
\label{steermeasurements2}\end{figure}

Now we evaluate the right hand side of \erf{SingleInequality4} for the case when
Alice conditions using inefficient photodetection. The quantities
in the right hand side of \erf{SingleInequality4} were already determined in Sec.~\ref{introphoto}.
Substituting these in with $\chi=1/2$, we find that \erf{SingleInequality4} will be {\em violated} iff
\beq
\sqrt{\frac{2}{\pi}}\,\eta\sqrt{\eta_{\rm h}} >
f(n)\sqrt{\eta(2-\eta-\eta\eta_{\rm p})}. \label{sufevenotinf}
\eeq
In the limit $n\to\infty$, this can be rearranged to give the simple inequality
\beq \eta > \frac{4}{2+\pi\eta_{\rm h}+2\eta_{\rm p}}. \label{sufeven}
\eeq
This is a {\em sufficient condition} on the three parameters
($\eta, \eta_{\rm h}, \eta_{\rm p}$) for experimental
demonstration of EPR-steering. It may be visualized by a
contour plot as shown in Fig.~\ref{steermeasurements2}.   The
contours show the required minimum value of $\eta$, as a function
of $\eta_{\rm h}$ and  $\eta_{\rm p}$. Note that for low measurement
efficiencies no contours are plotted as the preparation efficiencies
required according to \erf{sufeven} would be unphysical ($\eta>1$).
Note also that even without photodetection ($\eta_{\rm p}=0$), it would be
possible (in principle) to satisfy this sufficient condition provided $\eta_{\rm h}$ and $\eta$
are high enough. But with $\eta_{\rm h} = 0$ the inequality can never be satisfied,
and this is because it is impossible for Alice to demonstrate EPR-steering with a single
measurement.  

\subsection{Unevenly split single photon state}\label{Uneven}

The above analysis generalizes easily to the case of an unevenly split
photon $W_\eta^\chi$,  where $\chi$ can take any value
between 0 and 1.  The specific cases reported in
\cite{BabBreLvoPRL04} were $\chi=0.5$ and $\chi=0.92$.
We predict that the EPR-steering inequality  \erf{SingleInequality4} can be violated if
\beq
2\sqrt{\frac{2}{\pi}}\eta\sqrt{\eta_{\rm h}\chi(1-\chi)} >
f(n)\sqrt{4\eta\chi\left[1-\eta\chi-\eta\eta_{\rm
p}(1-\chi)\right]}. \eeq
Retaining this time a fully general result by not taking the number of Alice's homodyne
settings to infinity, we find the following sufficient condition:
 \beq
\eta  \left[ \chi +  (1-\chi)\left(\eta_{\rm p}+\frac{2}{\pi}
\frac{\eta_{\rm h}}{[f(n)]^2}\right) \right] > 1.\label{sufficient}
\eeq 
In   Sec.~\ref{Comparison} we give examples showing when \erf{sufficient} with finite settings could be satisfied in a real experiment.

\section{Necessary condition  for EPR-steering}\label{NecCond}

While the sufficient condition of \erf{sufficient} is a
useful guide to experiments, failure to satisfy this condition
(in the limit $n\to \infty$) does not mean that it would be impossible to demonstrate EPR-steering
with the state $W_\eta^\chi$ using homodyne detection (with arbitrary phase)
and photon counting, with
efficiencies $\eta_{\rm h}$ and $\eta_{\rm p}$ respectively. This is because
there may exist a better EPR-steering inequality than the one
(\ref{SingleInequality4}) we have derived --- that is, an inequality that can be violated
in a larger region of parameter space.
Perhaps surprisingly, however, we can derive a {\em necessary} condition
on the experimental parameters $\chi$, $\eta$, $\eta_{\rm h}$ and $\eta_{\rm p}$
for EPR-steering to be demonstrated, that makes no assumption on the
inequality to be tested.

First, we consider the basic fact that in order for EPR-steering to
take place, Alice must be able to perform at least two distinct measurements.
In an experimental setup, if an imperfectly prepared (with
efficiency $\eta$) single photon is mixed with the vacuum at a
$\chi : (1-\chi)$ beam splitter, then we arrive at the situation
where Alice and Bob share the state $W_\eta^\chi$.  In such an
experiment, Alice is assumed to receive on average
a proportion $(1-\chi)\eta$ of the light from the initial single photon. 
We  now consider what Alice could do with this light. 

\begin{figure}[]
\begin{center}
\includegraphics[width=7.5cm]{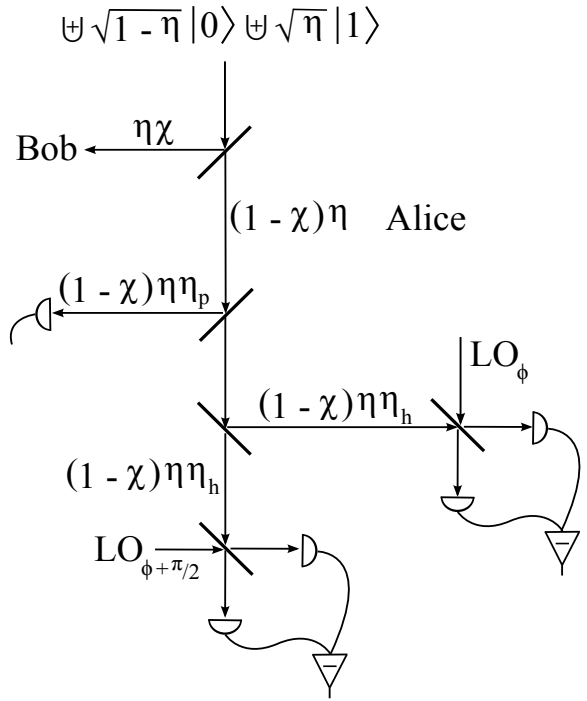}
\end{center}
\caption[Possible measurement scheme for Alice]{If \erf{necess}  is
satisfied  then it is possible that Alice could use the
measurement scheme depicted here, which would \emph{not}
constitute EPR-steering.  That is,  providing she had arbitrarily good 
detectors, 
she could partition her fraction $(1-\chi)\eta$ of
the light and use $\eta_{\rm p}$ of it for performing 
photodetection,  and two portions of $\eta_{\rm h}$ to
perform homodyne measurements of orthogonal  quadratures.   This is
essentially a \emph{single}, though complicated, measurement
scheme which cannot demonstrate EPR-steering.}
\label{necessFig}\end{figure}

 It is 
central to the definition of EPR-steering that Bob cannot trust 
Alice. Thus he should not simply believe her if she says that her photodetectors and homodyne photo-receivers
have efficiency $\eta_{\rm p}$ and $\eta_{\rm h}$ respectively. 
It could be that she actually has perfect detectors. Let us assume that to be the case. 
It follows that if the
following inequality \beq (1-\chi)\eta \eta_{\rm p} +
(1-\chi)\eta 2\eta_{\rm h} \leq (1-\chi)\eta, \label{necess} \eeq is
satisfied then,  Alice 
could use a complicated scheme to partition her fraction of the
light, and simultaneously perform photodetection and homodyne
measurements of two orthogonal phases, with efficiencies $\eta_{\rm p}$,
$\eta_{\rm h}$, and $\eta_{\rm h}$. 
This is demonstrated in Fig.~\ref{necessFig} (for the case where the inequality is saturated), 
by showing the proportion of the original single photon which Alice is able to send to each of her 
(assumed perfect) detectors.   Now homodyne
measurements of two orthogonal phases with efficiency $\eta_{\rm h}$
enables Alice to simulate the result of a homodyne measurement at {\em any}
phase $\theta$, by suitably combining the two results, with the same efficiency $\eta_{\rm h}$.
Thus this setup allows Alice to perform {\em all} of her possible measurements
in a single measurement (i.e. with no change of the apparatus). By definition however, a single measurement
by Alice cannot demonstrate EPR-steering.  Therefore, it is necessary
that \erf{necess} be {\em violated} in order for EPR-steering to be
possible.

\begin{figure}[]
\begin{center}
\includegraphics[width=7.5cm]{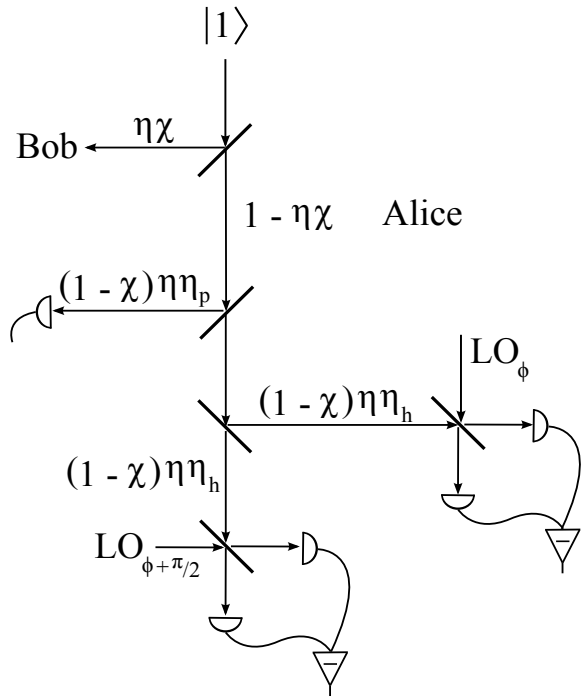}
\end{center}
\caption[Necessary condition]{The setup is the same as in Fig.~\ref{necessFig} however now Bob does not trust the initial preparation of the shared state. He knows that on average he receives $\eta\chi$ of the single photon, but he now assumes that Alice has access to the remaining $1-\eta\chi$ of the initial state,  as well as having arbitrarily good detectors. In this case,  she could again use a complicated measurement scheme to partition her fraction of the initial state to simultaneously perform both photodetection and homodyne measurements of orthogonal phases with efficiencies $\eta_p$ and $\eta_h$ respectively provided that \erf{necess2} is satisfied.}\label{necessFig6}\end{figure}

It is obvious that \erf{necess} reduces to the much simpler condition $\eta_p+2\eta_h \leq 1$. 
 We write the condition as 
in \erf{necess} to more easily relate conceptually to the next inequality we derive, which is not as straightforward, 
and which is a stronger inequality.  Recall that a non-trusting
Bob is the concept at the heart of the definition of EPR-steering. Such a Bob 
would know that a clever Alice could obtain a larger fraction
of the light from the initial single photon than the $(1-\chi)\eta$ shown in Fig.~\ref{necessFig}. 
Rather, a devious Alice may obtain access  to the preparation of the initial single photon and
recover the fraction $1-\eta$ of the light from this state that
was thought to be lost in the inefficient preparation so that she receives a total fraction $1 - \eta\chi$. In this case 
 (and again assuming perfect detectors),  Alice 
can simultaneously obtain results for all of her possible measurements,
with the right efficiencies, by a complicated measurement on
her `boosted' fraction of the light, provided that 
\beq (1-\chi)\eta \eta_{\rm p} +
(1-\chi)\eta 2\eta_{\rm h} \leq  1-\chi\eta. \label{necess2} \eeq
This is demonstrated in Fig.~\ref{necessFig6}, for the 
case where the inequality is saturated and Alice must use every bit of light available. 
Thus, in order for  it to be possible for Bob to be convinced that Alice is not simply
performing a complicated single measurement on her fraction of the
light, the inequality (\ref{necess2}) must be violated.

Rearranging \erf{necess2}, we
arrive at a necessary condition for it to be possible to demonstrate EPR-steering
using the split-photon states $W_\eta^\chi$ and homodyne detection  and photon counting:
\beq \eta \left[\chi + (1-\chi) \ro{\eta_{\rm p} +2\eta_{\rm h} }
\right] >  1. \label{necessary} \eeq
Note the similarity in form to the sufficient condition (\ref{sufficient}), and note
that the necessary condition is strictly weaker  than the sufficient condition.

\section{Comparison with experiment}\label{Comparison}

\subsection{Necessary conditions}

Finally, we are in a position to reconsider the experiments
\cite{BabBreLvoPRL04,BabEtalPRL04}.
In the latter, Lvovsky and co-workers obtained experimental
parameters of $\eta=0.64$, and $\eta_{\rm h}=0.86$, and
in the former they used $\chi=0.5$ and $\chi = 0.92$.
 First let us
test these parameters in \erf{necessary} to determine if EPR-steering
was possible in this experiment.  Evaluating the left hand side of
the inequality we obtain $0.87$ and $0.68$ for $\chi=0.5$ and
$0.92$ respectively.  Clearly the necessary condition for
steerability is not satisfied under either of these circumstances.
Even if
we add photodetection, with an efficiency  $\eta_{\rm p}=0.3$
(which is experimentally feasible \cite{AltEtalOE05}), we only raise these figures
to $0.97$ and $0.69$ respectively, still short of the required value of $1$.

The easiest parameter to alter to try to improve the performance of
the experiment is the splitting ratio $\chi$.
As stated, the experiment of \cite{BabBreLvoPRL04} made use of a symmetric as
well as a decidedly asymmetric arrangement (with Bob obtaining a
much larger fraction of the light).  In order to facilitate the demonstration of
EPR-steering, we thus determine the optimal $\chi$.  In the example
above, the symmetric arrangement came much closer to satisfying
the necessary condition than the asymmetric one.  This might tempt
one to conclude that the symmetric situation is most
useful for a demonstration of EPR-steering.  In fact, this is
 true neither for the necessary  nor the sufficient criteria for EPR-steering.

Considering both \erf{sufficient} and \erf{necessary} it is
straightforward to see that $\chi=0$ is the optimal value for
satisfying these conditions.  However, physically this corresponds
to the asymmetric case of Alice obtaining all of the light, in
which case Alice and Bob do not share an entangled state at all
and EPR-steering cannot occur.  Thus, in practise the optimal
arrangement would be to choose $\chi \ll 1$. (Experimental imperfections
not modelled in our theory would presumably imply an optimal, small, value for $\chi$.)
That is, in the Babichev \ea
experiment \cite{BabBreLvoPRL04}, the asymmetry was weighted in the wrong direction ---
they had $\chi=0.92$, which sends almost all of the light to Bob.

We can understand this result intuitively, as it is Alice's detection efficiencies
that matter in the experiment, not Bob's. Providing Alice with \emph{less} of the initial state
makes it more difficult for her to influence (steer) Bob's
part of the state. If the setup used in \cite{BabBreLvoPRL04} were to be reversed
so that $\chi=0.08$, then with $\eta=0.64$, $\eta_{\rm h}=0.86$, and
no photodetection ($\eta_{\rm p}=0$), the left hand side of \erf{necessary}
evaluates to $1.06$, suggesting that  EPR-steering might be possible in an experiment similar to
that in Ref.~\cite{BabBreLvoPRL04}.
Including photodection with $\eta_{\rm p}=0.3$, gives a left hand side of $1.24$,
suggesting that EPR-steering {\em should} be possible using such an
enhanced experiment.

\subsection{Sufficient conditions}

In order to determine if these hypothetical experiments could
demonstrate EPR-steering using the inequalities we have derived,
 we need to test the sufficient condition
\erf{sufficient}.  We use the parameters of \cite{BabEtalPRL04} as
above, with $\chi=0.08$ and we assume eight homodyne measurement
settings. Using eight settings is few enough to seem
experimentally feasible, but a large enough number so that
$f(8)\simeq 0.641$ is not far from $f(\infty) =  2/\pi \simeq 0.637$.

With no photodetection, the left hand side of the
sufficient condition (\ref{sufficient}) evaluates to $0.84$. That is, unfortunately,
the EPR-steering inequality we derived would be a long way from being violated.
But including photodection with $\eta_{\rm p}=0.3$, gives a left hand side of $1.01$,
implying that it would just be possible to violate \erf{SingleInequality4}, and
so demonstrate EPR-steering, in this enhanced experiment.

In practice of course it would be
extremely difficult for experimentalists to satisfactorily observe
a violation which equates to $1.01 > 1.00$.  Thus, in order to
conclusively demonstrate EPR-steering of a single photon it
would be desirable to have a larger violation of the
steering criterion.  The appeal of our approach  is that we have provided a number of experimental
parameters that may be adjusted to facilitate EPR-steering.

Consider first the case where there is no photodetection, as in the original experiments.
Then if one chose $\chi = 0.05$, and if one
were able to improve the parameters to $\eta = 0.78$ and $\eta_{\rm h}=0.92$,
the sufficient condition (\ref{sufficient}) would be satisfied as follows: $1.10 > 1$.
Alternatively, with photodetection included with efficiency $\eta_{\rm p} = 0.30$, and with
 $\eta=0.66$ and $\eta_{\rm h}=0.90$, one would also find the same degree of violation. 
  If one had access to a more efficient single-photon detector, the requirements on $\eta$
 and $\eta_{\rm h}$ set by \erf{sufficient} become even less stringent. 
 Thus demonstrating a substantial violation of an EPR-steering inequality with a single
 photon could be achieved with only moderate improvements to experimental
techniques.

\section{Summary}\label{Summary}

In this work we have considered in detail the experimental prospects
for demonstrating quantum nonlocality with a single photon, using homodyne
detection. This question is of considerable interest
experimentally, with both Bell-nonlocality (violation of a CHSH-inequality, with post-selection) \cite{BabEtalPRL04},
and ``remote state preparation'' (i.e. the EPR-steering phenomenon) \cite{BabBreLvoPRL04}
of a single photon being addressed.
Our analysis here shows that while the impure state produced
in these experiments could not possibly be used to demonstrate violation of a CHSH-inequality
{\em without} post-selection, it could be used to demonstrate EPR-steering,
according to the definition established in Refs.~\cite{WisEtalPRL07,JonWisDohPRA07}, 
without post-selection.
Also, we showed that even given the efficiency of the homodyne detection used
in these experiments, it might still be possible to do a rigorous demonstration of EPR-steering,
if the asymmetry of the photon splitting were reversed from that used in Ref.~\cite{BabBreLvoPRL04}.

To actually demonstrate EPR-steering would require violating an EPR-steering
inequality, as defined in Ref.~\cite{CavEtalPRA09}. Here we have introduced a family of such
inequalities for homodyne detection on a single split photon, with an arbitrary number
 of different phase settings. Unfortunately with the experimental efficiency
of homodyne detection achieved in Ref.~\cite{BabEtalPRL04}, it would not be possible
to violate any of these inequalities. However, we also generalized our inequality by supplementing
homodyne detection with a photon detector. We showed
that with a realistic photon detector efficiency, with relatively few different homodyne phases,
and with only modest improvements to the efficiency of photon preparation and homodyne detection,
it should be possible to achieve a substantial violation of our inequality. Thus,
for rigorously demonstrating nonlocality (in the EPR-steering sense) of a single photon, in an
experiment in the near future, the prospects are good.

\acknowledgments 
HMW acknowledges useful discussions with A. Lovovsky. This research was conducted by the Australian Research Council Centre of Excellence for Quantum Computation and Communication Technology (project number CE110001029)

\appendix\label{app}
\section{Convexity proof}
We consider the function:
\beq
f(\{ \langle \hat{\sigma}_\theta \rangle\},\langle \hat{\sigma}_z\rangle)=\frac{1}{\pi}\int_{-\pi/2}^{\pi/2}d\theta a_\theta\langle \hat{\sigma}_\theta\rangle - \frac{2}{\pi}\sqrt{1-\langle \hat{\sigma}_z\rangle^2},\label{fconvex:app}
\eeq
with the aim of showing that
\begin{enumerate}
\item $f$ is a convex function of its arguments,
\item and that $f\leq 0$ $\forall\ a_\theta\in\{ -1,1\},\forall \rho$.
\end{enumerate}

In order to prove point 1, we must show that both terms in \erf{fconvex:app} are convex functions, as the sum of two convex functions is also a convex function \cite{RocCA70}.  The first term is trivially convex, as the integral is just the continuous limit of adding the arguments, which are linear, and hence convex. For the second term, we need simply examine a plot of $-\sqrt{1-z^2}$ for $z\in\{-1,1\}$ to verify that it has a convex (i.e. concave up) shape.

Now to prove point 2, we must show that $f \leq 0$,  $\forall\ a_\theta\in\{ -1,1\},\forall \rho$.  This amounts to showing that
\beq
\frac{1}{\pi}\int_{-\pi/2}^{\pi/2}d\theta a_\theta\langle \hat{\sigma}_\theta\rangle \leq \frac{2}{\pi}\sqrt{1-\langle \hat{\sigma}_z\rangle^2}.\label{fneg:app3}
\eeq  It was shown in Sec.~\ref{SingleInfiniteCriteria}, that the maximum of the left hand side is occurs when $a_\theta \equiv 1$.  Making this substitution and evaluating the integral means that we are required to prove the condition
\beq
\frac{2}{\pi}\langle \hat{\sigma}_x\rangle \leq \frac{2}{\pi}\sqrt{1-\langle \hat{\sigma}_z\rangle^2}.
\eeq  That this holds for all $\rho$ follows immediately from the condition that  $\langle \hat{\sigma}_x\rangle^2 +\langle \hat{\sigma}_z\rangle^2 \leq 1$.

\bibliographystyle{prsty}
\bibliography{paperbib_120311}

\end{document}